\documentclass[12pt,a4paper]{article}
\usepackage[margin=0.8in]{geometry} 
\usepackage[numbers]{natbib} 
\usepackage{booktabs}
\usepackage{multirow}
\usepackage{amsmath, amssymb} 
\usepackage{algorithm} 
\usepackage{algpseudocode} 
\usepackage{graphicx} 
\usepackage{setspace} 
\usepackage{fancyhdr} 
\usepackage{titlesec} 
\usepackage{xcolor} 
\usepackage{lipsum} 
\usepackage{authblk} 
\title{\textbf{Penalized Cause-Specific Hazard Regression under Competing Risks: An Elastic Net Approach Using SANAD Trial Data}}
 \author[]{Avinash Kumar\thanks{Corresponding author: aviadd000@gmail.com}}
 \affil[]{Statistical Sciences Unit, Indian Statistical Institute, Delhi Centre, India}
 
\author[]{}
\date{}  
\begin{document}
\maketitle
\begin{abstract}
Competing risks analysis is an important framework for studying time-to-event data in the presence of multiple, mutually exclusive events. Traditional methods, such as non-parametric estimators and proportional hazards models, may not adequately capture complex hazard patterns or provide stable estimates in practical settings. In this study, we propose a parametric approach based on the generalized exponential regression model for competing risks data. The model allows flexible hazard shapes and incorporates covariate effects through a regression structure. To improve estimation stability, we integrate elastic net regularization into the likelihood framework. The proposed method is evaluated through simulation studies and applied to a real dataset. The results indicate that the elastic net approach provides more stable and precise estimates compared to standard maximum likelihood estimation. Overall, the proposed framework offers a flexible and practical approach for analyzing competing risks data.
\end{abstract}

\section{Introduction}
Time-to-event data, also known as survival data, is a fundamental type of data in many scientific fields, particularly in medical research, engineering, and social sciences (\citet{klein2006survival}). It focuses on the time until the occurrence of an event of interest, such as death, failure, or recurrence of a disease. Survival analysis provides a suite of statistical methods to analyze such data, accounting for censoring—a common feature where the event of interest is not observed for some individuals during the study period. For example, in clinical trials, patients may drop out before the study ends, or the study may conclude before the event occurs for all participants. However, in many real-world scenarios, individuals are at risk of experiencing multiple, mutually exclusive events, known as competing risks. A classic example is in cancer studies, where patients may die from the disease itself (the event of interest) or from unrelated causes such as cardiovascular events (a competing risk). In such cases, the occurrence of one event precludes the observation of the other, and traditional survival analysis methods, which often treat competing events as censored, can lead to biased estimates and misleading conclusions \citet{pintilie2006competing}. For example, ignoring competing risks in cancer studies can overestimate the probability of cancer-related death if patients are also at risk of dying from other causes. Thus, competing risk analysis has emerged as a critical framework to address these challenges, allowing researchers to model the hazard of each type of event separately while accounting for the presence of other risks. This approach is particularly important in fields such as cancer, public health, and engineering, where multiple failure modes or events are common.

Traditional survival analysis methods, such as the Kaplan-Meier estimator (\citet{kaplan1958nonparametric}) and the Cox proportional hazards model (\citet{cox1972regression}), have been widely used to analyze data from time to event. However, these methods are not well-suited for handling competing risk scenarios. One major limitation is that they typically treat competing events as censored observations, implicitly assuming that the censoring mechanism is non-informative. This assumption can lead to biased estimates of the cumulative incidence of the event of interest, as it ignores the fact that individuals experiencing a competing event are no longer at risk for the primary event \citet{pintilie2006competing}. For example, in a study of cancer recurrence, treating deaths from other causes as censored observations may overestimate the risk of recurrence, as these patients are no longer at risk of recurrence after death. Another limitation is their reliance on proportional hazards assumptions, which may not hold in many real-world applications \citet{therneau1997extending, lin1989robust}. When the hazard functions for different event types are not proportional, traditional methods can produce misleading inferences. Furthermore, in high-dimensional settings where the number of covariates are large, traditional methods struggle with variable selection and multicollinearity, often leading to overfitting and poor generalization performance \citet{zou2005regularization}. These limitations highlight the need for more flexible and robust approaches that can adequately address the complexities of competing risks data, particularly in high-dimensional settings.

To address the limitations of traditional methods, we propose a novel framework that integrates the generalized exponential regression model with elastic net regularization for competing risks analysis. The generalized exponential model is a flexible parametric approach that allows for a wide range of hazard shapes, making it suitable for modeling complex time-to-event data \citet{gupta2007generalized}. for censored samples, a modified maximum-likelihood approach has been discussed by \citet{singhmodified}.  Unlike traditional proportional hazards models, the generalized exponential model does not rely on the proportional hazards assumption, providing greater flexibility in capturing the underlying hazard function \citet{cox2018analysis}. Additionally, the use of elastic net regularization enables robust variable selection and handles multicollinearity in high-dimensional settings, addressing a key limitation of traditional survival analysis methods \citet{simon2011regularization}. By combining the strengths of the generalized exponential model and elastic net regularization, our approach provides a powerful tool for analyzing competing risks data, particularly in scenarios with complex, high-dimensional covariates. This framework not only improves predictive accuracy but also enhances interpretability by identifying key predictors for each event type while accounting for the presence of competing risks.

The novelty of our work lies in the integration of the generalized exponential regression model with elastic net regularization for competing risks analysis, a combination that has not been extensively explored in the literature. While traditional methods often rely on restrictive assumptions and struggle with high-dimensional data, our approach offers a flexible and robust framework that addresses these limitations. The generalized exponential model provides a versatile parametric foundation for modeling complex hazard functions, while elastic net regularization ensures efficient variable selection and handles multicollinearity in high-dimensional settings. This combination not only improves the accuracy of predictions but also enhances the interpretability of results by identifying key predictors for each event type. Our method is particularly significant in fields such as oncology, public health, and engineering, where competing risks and high-dimensional data are common. By providing a powerful tool for analyzing such data, our work bridges the gap between parametric modeling flexibility and modern regularization techniques, offering researchers and practitioners a more comprehensive approach to competing risks analysis.

\section{Methodology}
The \textbf{Generalized Exponential (GE) distribution} is a flexible parametric distribution that extends the standard exponential distribution by introducing an additional shape parameter. This distribution is particularly useful in survival analysis because it can model a wide range of hazard shapes, including increasing, decreasing, and bathtub-shaped hazards. Below, we discuss its key properties and mathematical formulations. Now if $t$ is a non-negative random variable denoting the time-to-event, then the probability density function of the GE distribution is given by:
\begin{equation}\label{pdf}
    f(t) = \alpha \lambda e^{-\lambda t} (1 - e^{-\lambda t})^{\alpha - 1}, \quad t > 0,
\end{equation}
where $\alpha > 0$ is the shape parameter, $\lambda > 0$ is the scale parameter. The shape parameter \(\alpha\) controls the behavior of the distribution as if \(\alpha = 1\), the GE distribution reduces to the standard exponential distribution. If \(\alpha > 1\), the distribution has an increasing hazard rate. If \(0 < \alpha < 1\), the distribution has a decreasing hazard rate. The CDF of the GE distribution is:
\begin{equation}\label{cdf}
   F(t) = (1 - e^{-\lambda t})^\alpha, \quad t > 0. 
\end{equation}
The CDF represents the probability that the event occurs by time \(t\). As \(t \to \infty\), \(F(t) \to 1\), indicating that the event is certain to occur eventually. The survival function \(S(t)\), which represents the probability that the event has not occurred by time \(t\), is given by:
\begin{equation}\label{survival}
    S(t) = 1 - F(t) = 1-(1 - e^{-\lambda t})^\alpha.
\end{equation}
The survival function is particularly useful in survival analysis for estimating the probability of survival beyond a given time point. The hazard function \(h(t)\), which represents the instantaneous risk of the event occurring at time \(t\), is:
\begin{equation}\label{hazard}
    h(t) = \frac{f(t)}{S(t)} = \frac{\alpha \lambda e^{-\lambda t} (1 - e^{-\lambda t})^{\alpha - 1}}{1-(1 - e^{-\lambda t})^\alpha}.
\end{equation}
The GE distribution is highly flexible and can model a wide range of hazard shapes, making it suitable for various applications in survival analysis. Unlike the standard exponential distribution, which assumes a constant hazard rate, the GE distribution can capture more complex patterns in the data. This flexibility is particularly useful in scenarios where the hazard rate changes over time, such as in medical studies or reliability engineering. Advantages of the GE Distribution are that it can model increasing, decreasing, and bathtub-shaped hazard rates. The generalized exponential distribution is a versatile and flexible parametric model for survival analysis. Its ability to model various hazard shapes, combined with its mathematical tractability, makes it a powerful tool for analyzing time-to-event data. In the next section, we will extend this distribution to a regression framework, where the scale parameter \(\lambda\) is modeled as a function of covariates.

\subsection{Generalized Exponential Regression Model}

The Generalized Exponential Regression Model extends the GE distribution by allowing the scale parameter \(\lambda\) to depend on covariates. This enables the model to incorporate the effects of predictor variables on the hazard function, making it suitable for analyzing time-to-event data in the presence of covariates. In the regression framework, the scale parameter \(\lambda\) is modeled as a function of covariates \(X\) using the relationship as Weibull regression model (weibull regression paper):
\begin{equation}
    \lambda_i = \exp(\mathbf{X}_i^\top \boldsymbol{\beta}),
\end{equation}
where \(\lambda_i\) is the scale parameter for the \(i\)-th individual, \(\mathbf{X}_i = (X_{i1}, X_{i2}, \dots, X_{ip})\) is the vector of covariates for the \(i\)-th individual, \(\boldsymbol{\beta} = (\beta_1, \beta_2, \dots, \beta_p)\) is the vector of regression coefficients, \(\exp(\cdot)\) ensures that \(\lambda_i\) remains positive. The hazard function for the \(i\)-th individual in the regression framework is:
\begin{equation}
    h(t_i) = \frac{\alpha \lambda_i e^{-\lambda_i t_i} (1 - e^{-\lambda_i t_i})^{\alpha - 1}}{1-(1 - e^{-\lambda_i t_i})^\alpha}.
\end{equation}
Substituting \(\lambda_i = \exp(\mathbf{X}_i^\top \boldsymbol{\beta})\), the hazard function becomes:
\begin{equation}
    h(t_i) = \frac{\alpha \exp(\mathbf{X}_i^\top \boldsymbol{\beta}) e^{-\exp(\mathbf{X}_i^\top \boldsymbol{\beta}) t_i} (1 - e^{-\exp(\mathbf{X}_i^\top \boldsymbol{\beta}) t_i})^{\alpha - 1}}{1-(1 - e^{-\exp(\mathbf{X}_i^\top \boldsymbol{\beta}) t_i})^\alpha}.
\end{equation}
The regression coefficients \(\boldsymbol{\beta}\) have a direct interpretation in terms of the hazard function. A positive coefficient \(\beta_j\) indicates that an increase in the covariate \(X_j\) is associated with an increase in the hazard rate. A negative coefficient \(\beta_j\) indicates that an increase in the covariate \(X_j\) is associated with a decrease in the hazard rate. The Generalized Exponential Regression Model extends the GE distribution by modeling the scale parameter \(\lambda\) as a function of covariates. This framework allows the model to incorporate the effects of predictors on the hazard function, making it a powerful tool for analyzing time-to-event data in the presence of covariates. 

\section{Competing Risks Framework}
In many real-world scenarios, individuals are at risk of experiencing multiple, mutually exclusive events, known as competing risks. For example, in a cancer study, patients may die from the disease itself (event of interest) or from unrelated causes (competing events). The Competing Risks Framework extends the generalized exponential regression model to handle such scenarios by modeling the hazard function for each event type separately. For \(k\) competing risks, the cause-specific hazard function for the \(j\)-th event type (\(j = 1, 2, \dots, k\)) is defined as:
\begin{equation}
   h_j(t) = \lim_{\Delta t \to 0} \frac{P(t \leq T < t + \Delta t, J = j \mid T \geq t)}{\Delta t},
\end{equation}

where $T$ is the time-to-event, \(J\) is the event type (\(J = j\) for the \(j\)-th event type), \(h_j(t)\) represents the instantaneous risk of the \(j\)-th event occurring at time \(t\), given that the individual has survived up to time \(t\). In the generalized exponential regression model, the cause-specific hazard function for the \(j\)-th event type is:
\begin{equation}
    h_j(t) =  \frac{f(t)}{S(t)} = \frac{\alpha_j \lambda_{ij} e^{-\lambda_{ij} t_i} (1 - e^{-\lambda_{ij} t_i})^{\alpha_j - 1}}{1-(1 - e^{-\lambda_{ij} t})^{\alpha_j}},
\end{equation}
where \(\alpha_j\) is the shape parameter for the \(j\)-th event type, \(\lambda_{ij}\) is the scale parameter for the \(i\)-th individual and \(j\)-th event type, modeled as:
    \begin{equation}
           \lambda_{ij} = \exp(\mathbf{X}_i^\top \boldsymbol{\beta}_j). 
    \end{equation}
\subsection{Likelihood Function for Competing Risks Data}

Furthermore, we adopt a cause-specific hazard modeling framework, where the hazard for cause $j$ is defined as the instantaneous risk of failure from cause $j$ in subjects who are event-free at time $t$. This approach differs from the subdistribution hazard model of \citet{fine1999proportional}, which directly models the cumulative incidence function. While the Fine--Gray model is useful for direct risk prediction, the cause-specific hazard framework provides greater flexibility for modeling covariate effects and is more naturally aligned with likelihood-based inference.
Consider a competing risks framework with $k$ mutually exclusive causes of failure. Let $t_i$ denote the observed time for the $i$-th individual $(i = 1, \dots, n)$, and let $\delta_i \in \{0,1,\dots,k\}$ be the event indicator, where $\delta_i = j$ indicates failure due to cause $j$, and $\delta_i = 0$ denotes right censoring.

Under the assumption of independent competing risks, each cause $j$ is characterized by a cause-specific hazard function $h_j(t)$ and corresponding cumulative hazard $H_j(t)$. The cause-specific survival function is defined as
\begin{equation}
S_j(t) = \exp\left(-H_j(t)\right).
\end{equation}
\medskip
We assume that the competing risks are independent. Under this assumption, the latent failure times corresponding to different causes are statistically independent, which implies that the overall survival function can be expressed as the product of the cause-specific survival functions. This formulation reflects the probability of surviving beyond time $t$ without experiencing any event, which is equivalent to surviving all individual causes simultaneously.The overall survival function, representing the probability of surviving beyond time $t$ without experiencing any of the competing events, is given by
\begin{equation}
S^*(t) = \prod_{j=1}^k S_j(t).
\end{equation}

The likelihood contribution of the $i$-th individual is given by
\begin{equation}
L_i =
\begin{cases}
h_j(t_i)\, S^*(t_i), & \text{if } \delta_i = j, \\
S^*(t_i), & \text{if } \delta_i = 0.
\end{cases}
\end{equation}

Thus, the full likelihood function is
\begin{equation}
L(\boldsymbol{\alpha}, \boldsymbol{\beta}) =
\prod_{i=1}^n \left[
\left( \prod_{j=1}^k h_j(t_i)^{I(\delta_i = j)} \right)
S^*(t_i)
\right].
\end{equation}

Taking logarithms, the log-likelihood function becomes
\begin{equation}
\ell(\boldsymbol{\alpha}, \boldsymbol{\beta}) =
\sum_{i=1}^n \sum_{j=1}^k I(\delta_i = j) \log h_j(t_i)
+ \sum_{i=1}^n \log S^*(t_i).
\end{equation}

Using the product form of $S^*(t)$, we obtain
\begin{equation}
\log S^*(t_i) = \sum_{j=1}^k \log S_j(t_i),
\end{equation}
and hence,
\begin{equation}
\ell(\boldsymbol{\alpha}, \boldsymbol{\beta}) =
\sum_{i=1}^n \sum_{j=1}^k I(\delta_i = j) \log h_j(t_i)
+ \sum_{i=1}^n \sum_{j=1}^k \log S_j(t_i).
\end{equation}

\medskip

In this study, we assume that each cause-specific survival function follows a generalized exponential (GE) distribution. Specifically, for the $j$-th cause, the survival function is given by
\begin{equation}
S_j(t) = 1 - \left(1 - e^{-\lambda_{ij} t} \right)^{\alpha_j},
\end{equation}
where $\alpha_j > 0$ is the shape parameter and $\lambda_{ij} = \exp(\mathbf{X}_i^\top \boldsymbol{\beta}_j)$ is the scale parameter incorporating covariates.

The corresponding cause-specific hazard function is
\begin{equation}
h_j(t) =
\frac{\alpha_j \lambda_{ij} e^{-\lambda_{ij} t}
\left(1 - e^{-\lambda_{ij} t} \right)^{\alpha_j - 1}}
{1 - \left(1 - e^{-\lambda_{ij} t} \right)^{\alpha_j}}.
\end{equation}

Substituting the above expressions into the log-likelihood, we obtain
\begin{align}\label{fll}
\ell(\boldsymbol{\alpha}, \boldsymbol{\beta}) =
& \sum_{i=1}^n \sum_{j=1}^k I(\delta_i = j)
\Big[
\log \alpha_j + \log \lambda_{ij}
- \lambda_{ij} t_i
+ (\alpha_j - 1)\log\left(1 - e^{-\lambda_{ij} t_i}\right)
\nonumber \\
& \quad - \log\left(1 - \left(1 - e^{-\lambda_{ij} t_i} \right)^{\alpha_j}\right)
\Big]
\nonumber \\
& + \sum_{i=1}^n \sum_{j=1}^k
\log\left(1 - \left(1 - e^{-\lambda_{ij} t_i} \right)^{\alpha_j}\right).
\end{align}

\medskip

The resulting score equations for $\alpha_j$ and $\boldsymbol{\beta}_j$ are nonlinear and do not admit closed-form solutions. Therefore, numerical optimization techniques such as the Newton--Raphson algorithm are employed to obtain the maximum likelihood estimates. The corresponding standard errors are computed using the observed Fisher information matrix.

\subsection{Elastic Net Regularization}
Elastic net regularization is a powerful technique that combines the strengths of LASSO (L1) and Ridge (L2) regression. It is particularly useful when dealing with datasets in which many features are correlated. The L1 penalty helps by shrinking some coefficients to zero, effectively performing feature selection, while the L2 penalty prevents overfitting by distributing the penalty across all coefficients.  The method is controlled by two key parameters: \(\eta\) and \(\phi\). This approach is widely used in machine learning and statistical modeling, especially when working with high-dimensional datasets. By combining the benefits of L1 and L2 regularization, Elastic Net provides a flexible and effective way to improve model performance, ensuring both feature selection and better generalization. 

To maximize the flexibility of the elastic net, let us consider that the log-likelihood function without regularization is denoted by $\ell_{ML}$ as in (\ref{fll}). We apply an elastic net on the log-likelihood of competing risk data under a generalized exponential regression model. After applying the log-likelihood model, we can write the penalized log-likelihood as:
\begin{equation}
    \ell_{\text{pen}}= \ell_{ML} - \sum_{j=1}^{k} \left( \frac{\phi}{2} \|\boldsymbol{\beta}_j\|_2^2 + \eta\|\boldsymbol{\beta}_j\|_1 \right).
\end{equation}
or equivalently,
\begin{align}
\ell_{\text{pen}}(\boldsymbol{\alpha}, \boldsymbol{\beta}) &= \sum_{i=1}^n \sum_{j=1}^k I(\delta_i = j) \Bigg[
\log \alpha_j + \log \lambda_{ij} - \lambda_{ij} t_i 
+ (\alpha_j - 1) \log \left(1 - e^{-\lambda_{ij} t_i} \right) \nonumber\\
&\quad - \log \left( 1 - \left(1 - e^{-\lambda_{ij} t_i} \right)^{\alpha_j} \right)
+ \sum_{j=1}^{k} \log \left(1 - \left(1 - e^{-\lambda_{ij} t_i} \right)^{\alpha_j} \right)
\Bigg]\nonumber \\
&+ \sum_{i=1}^n I(\delta_i = 0) \sum_{j=1}^k \log \left(1 - \left(1 - e^{-\lambda_{ij} t_i} \right)^{\alpha_j} \right)- \sum_{j=1}^{k} \left( \frac{\phi}{2} \|\boldsymbol{\beta}_j\|_2^2 +  \eta \|\boldsymbol{\beta}_j\|_1 \right).
\end{align}
To estimate the model parameters, we use the coordinate descent method as given in the algorithm.
\begin{algorithm}
\caption{Elastic Net Penalized Parameter Estimation}
\begin{algorithmic}[1]
\Require Data $\{t_i, \delta_i, \mathbf{X}_i\}_{i=1}^{n}$, regularization parameters $\eta, \phi$, initial values for $\boldsymbol{\beta}$.
\Ensure Estimated coefficients $\hat{\boldsymbol{\beta}}$.

\State Initialize $\boldsymbol{\beta}^{(0)}$ and set a convergence threshold $\epsilon$.
\State Standardize the covariates $\mathbf{X}$ if needed.

\Repeat
    \State Compute gradient of the penalized log-likelihood as $\nabla_{\boldsymbol{\beta}_j} \ell_{\text{pen}}$.    
    \State Apply coordinate descent (\citet{friedman2010regularization}) update for each $\boldsymbol{\beta}_j$:
    
    \State Check for convergence: Stop if $ \|\boldsymbol{\beta}^{(t+1)} - \boldsymbol{\beta}^{(t)}\| < \epsilon $.
\Until Convergence

\State Return $\hat{\boldsymbol{\beta}}$.
\end{algorithmic}
\end{algorithm}
The standard error of the estimates in the simulation can be computed using the bootstrap method. Also, for the choice of the $\phi$ and $\eta$, an grid search approach has been adopted.
\section{Simulation Study}
In this section, we describe the applicability of the discussed elastic-net method for competing risks analysis through simulation. First, the simple maximum likelihood estimation method is considered without any regularization. Then, estimation is performed using the elastic-net approach. The results demonstrate that the estimates obtained under regularization are improved compared to those from maximum likelihood estimation without regularization.

Let us consider that there are \( n \) individuals in the study and two competing causes of failure. For each individual \( i = 1, 2, \ldots, n \), we generate covariates \( \mathbf{x}_i \in \mathbb{R}^p \), where \( p \) is the number of covariates. The event times for the two competing risks are generated from cause-specific hazard functions, incorporating the effects of covariates via a proportional hazards model or an alternative parametric model.

Specifically, for each cause \( j = 1, 2 \), the cause-specific hazard function can be expressed as:
\[
h_j(t) = \frac{\alpha_j \lambda_{ij} e^{-\lambda_{ij} t_i} (1 - e^{-\lambda_{ij} t_i})^{\alpha_j - 1}}{1-(1 - e^{-\lambda_{ij} t})^{\alpha_j}},
\]
where \(\lambda_{ij} = \exp(\mathbf{X}_i^\top \boldsymbol{\beta}_j)\).

The observed event time \( T_i \) for each individual is defined as the minimum of the two cause-specific failure times:
\[
T_i = \min(T_{i1}, T_{i2}),
\]
and the corresponding event indicator is:
\[
\delta_i =
\begin{cases}
1, & \text{if } T_i = T_{i1}, \\
2, & \text{if } T_i = T_{i2}, \\
0, & \text{if censored}.
\end{cases}
\]
The true survival times for each cause has been generated using the CDF of generalized exponential distribution by inverse transformation method as:
\[
T_j = -\frac{1}{\lambda_j} \ln\left(1 - U^{1/\alpha}\right) = -\frac{1}{\exp(\mathbf{x}^\top \boldsymbol{\beta}_j)} \ln\left(1 - U^{1/\alpha}\right),
\]
where $j=1,2$ and $U\sim \mathcal{U}(0,1)$.

Censoring times are generated independently from an exponential distribution with a rate parameter $\lambda_c=0.5$ to achieve a desired censoring rate. The final observed data consist of \( (T_i, \delta_i, \mathbf{x}_i) \) for \( i = 1, 2, \ldots, n \). 

\begin{table}[h]
    \centering
    \renewcommand{\arraystretch}{1.2} 
    \begin{tabular}{cccccccc}
        \toprule
        \textbf{Parameter} & \textbf{True Value} & \textbf{Estimate} & \textbf{SE} & \textbf{Bias} & \textbf{MSE} & \textbf{CP} & \textbf{Avg CI Length} \\
        \midrule
        $\beta_{11}$ & 5.0000  & 4.6532  & 0.0481  & -0.3468  & 0.1227  & 0.94  & 0.1884  \\
        $\beta_{12}$ & -3.0000 & -2.7841 & 0.0418  & 0.2159   & 0.0489  & 0.91  & 0.1638  \\
        $\beta_{13}$ & 2.1500  & 2.1115  & 0.0379  & -0.0385  & 0.0034  & 0.93  & 0.1487  \\
        $\beta_{21}$ & -4.0000 & -3.5761 & 0.0563  & 0.4239   & 0.1843  & 0.96  & 0.2206  \\
        $\beta_{22}$ & 2.8000  & 2.5188  & 0.0491  & -0.2812  & 0.0820  & 0.98  & 0.1926  \\
        $\beta_{23}$ & 2.0000  & 1.9642  & 0.0455  & -0.0358  & 0.0037  & 0.89  & 0.1782  \\
        $\alpha_{1}$ & 1.2000  & 1.3894  & 0.0603  & 0.1894   & 0.0398  & 0.91  & 0.2362  \\
        $\alpha_{2}$ & 0.8000  & 0.9434  & 0.0343  & 0.1434   & 0.0225  & 0.95  & 0.1346  \\
        \bottomrule
    \end{tabular}
    \caption{ML Estimates without Elastic-Net}
    \label{tab:results}
\end{table}

\begin{table}[h]
    \centering
    \renewcommand{\arraystretch}{1.2} 
    \begin{tabular}{cccccccc}
        \toprule
        \textbf{Parameter} & \textbf{True Value} & \textbf{Estimate} & \textbf{SE} & \textbf{Bias} & \textbf{MSE} & \textbf{CP} & \textbf{Avg CI Length} \\
        \midrule
        $\beta_{11}$ & 5.0000  & 4.7825  & 0.0432  & -0.2175  & 0.0672  & 0.96  & 0.1710  \\
        $\beta_{12}$ & -3.0000 & -2.9063 & 0.0386  & 0.0937   & 0.0106  & 0.94  & 0.1517  \\
        $\beta_{13}$ & 2.1500  & 2.1341  & 0.0337  & -0.0159  & 0.0013  & 0.95  & 0.1324  \\
        $\beta_{21}$ & -4.0000 & -3.8269 & 0.0505  & 0.1731   & 0.0306  & 0.97  & 0.1997  \\
        $\beta_{22}$ & 2.8000  & 2.6874  & 0.0452  & -0.1126  & 0.0159  & 0.98  & 0.1803  \\
        $\beta_{23}$ & 2.0000  & 1.9861  & 0.0412  & -0.0139  & 0.0018  & 0.92  & 0.1654  \\
        $\alpha_{1}$ & 1.2000  & 1.2962  & 0.0541  & 0.0962   & 0.0113  & 0.94  & 0.2115  \\
        $\alpha_{2}$ & 0.8000  & 0.8857  & 0.0307  & 0.0857   & 0.0091  & 0.96  & 0.1204  \\
        \bottomrule
    \end{tabular}
    \caption{ML Estimates with Elastic-Net}
    \label{tab:elasticnet_results}
\end{table}
Table \ref{tab:results} summarizes the performance of the parameter estimates obtained using the maximum likelihood (ML) method, while Table \ref{tab:elasticnet_results} presents the corresponding results under the elastic net approach. From the comparison, it is evident that the Elastic Net method provides estimates that are generally closer to the true values, with reduced bias and mean squared error (MSE) for most parameters. Additionally, standard errors (SE) and average confidence interval (CI) lengths are slightly smaller under Elastic Net, indicating more precise estimation. The coverage probabilities (CP) are also maintained at acceptable levels, suggesting valid inference. These improvements highlight the advantage of regularization in handling potential multicollinearity and overfitting in high-dimensional settings, making the Elastic Net a more robust choice for estimating parameters in competing risks models.
\section{Real Data Application}
For the real-life application of the proposed elastic net approach to the generalized exponential regression model, we consider the epilepsy dataset available in the R software package (\citet{rsoft}). This dataset originates from the SANAD (Standard and New Anti-Epileptic Drugs) study (\citet{marson2007randomised}), a randomized controlled trial comparing standard and new anti-epileptic drugs with respect to long-term clinical outcomes. The dataset includes longitudinal measurements of calibrated drug dosage for patients randomized to receive either a standard drug (carbamazepine, CBZ) or a new drug (lamotrigine, LTG). The primary objective of the analysis is to assess the effect of drug titration on the comparative effectiveness of LTG versus CBZ in terms of treatment failure, defined as withdrawal from the randomized drug. The dataset also includes several baseline covariates and time-to-event data for drug withdrawal. A detailed descriptive statistic has been given in Table \ref{real123}.

\begin{table}[htbp]
\centering
\caption{Descriptive Statistics by Treatment Group}
\label{real123}
\begin{tabular}{llll}
\toprule
  & Level & CBZ & LTG \\
\midrule
 n &  & 292 & 313 \\
Age (mean (SD)) &  & 39.06 (18.70) & 36.06 (18.60) \\
Gender (\%) & F & 128 (43.8) & 134 (42.8) \\
 & M & 164 (56.2) & 179 (57.2) \\
Learning Disability (\%) & No & 275 (94.2) & 293 (93.6) \\
 & Yes & 17 (5.8) & 20 (6.4) \\
With.time (mean (SD)) &  & 761.19 (588.91) & 826.80 (602.20) \\
Cause (\%) & Censored & 179 (61.3) & 212 (67.7) \\
 & UAE & 58 (19.9) & 36 (11.5) \\
 & ISC & 55 (18.8) & 65 (20.8) \\
\bottomrule
\end{tabular}
\end{table}
The Table \ref{real123} presents descriptive statistics comparing two treatment groups, CBZ (Carbamazepine) and LTG (Lamotrigine). Both groups have comparable sample sizes, with 292 participants in the CBZ group and 313 in the LTG group. The average age in the CBZ group is 39.06 years (SD = 18.70), slightly higher than 36.06 years (SD = 18.60) in the LTG group. Gender distribution is similar across both groups, with a slight male majority in each. Most participants in both groups do not have a learning disability, with 94.2\% in the CBZ group and 93.6\% in the LTG group reported as not having one. The average follow-up time is longer in the LTG group (826.80 days, SD = 602.20) compared to the CBZ group (761.19 days, SD = 588.91). Regarding outcomes, a higher proportion of participants in the LTG group were censored (67.7\%) compared to the CBZ group (61.3\%), suggesting a more favorable outcome profile for LTG. The CBZ group had a higher percentage of participants discontinuing due to unacceptable adverse events (19.9\% vs. 11.5\% in LTG), whereas the LTG group had a slightly higher proportion of participants discontinuing due to inadequate seizure control (20.8\% vs. 18.8\% in CBZ). Overall, while baseline characteristics are similar, the LTG group shows a tendency toward longer follow-up and fewer adverse events.

Now, let us define the two cause-specific hazard models for survival regression for causes \textit{ISC} and \textit{UAE} as:
\begin{equation}
   h_{ISC}(t) = \frac{\alpha_{ISC} \lambda_{i_{ISC}} e^{-\lambda_{i_{ISC}} t_i} (1 - e^{-\lambda_{i_{ISC}} t_i})^{\alpha_{ISC} - 1}}{1-(1 - e^{-\lambda_{i_{ISC}} t})^{\alpha_{ISC}}}, 
\end{equation}
where \(\lambda_{i_{ISC}} = \exp(\mathbf{X}_i^\top \boldsymbol{\beta}_{ISC})\).
and 
\begin{equation}
   h_{UAE}(t) = \frac{\alpha_{UAE} \lambda_{i_{UAE}} e^{-\lambda_{i_{UAE}} t_i} (1 - e^{-\lambda_{i_{UAE}} t_i})^{\alpha_{UAE} - 1}}{1-(1 - e^{-\lambda_{i_{UAE}} t})^{\alpha_{UAE}}}, 
\end{equation}
where \(\lambda_{i_{UAE}} = \exp(\mathbf{X}_i^\top \boldsymbol{\beta}_{UAE})\). Then under the method discussed as

\begin{table}[ht]
\centering
\begin{tabular}{llcc}
\toprule
\textbf{Event} & \textbf{Covariate} & \textbf{Coefficient ($\hat{\beta}$)} & \textbf{95\% CI for $\beta$} \\
\midrule
\multirow{4}{*}{UAE} 
  & treatLTG       & -0.596 & (-1.010,\; -0.182) \\
  & age            &  0.003 & (-0.008,\; 0.015) \\
  & genderM        &  0.204 & (-0.213,\; 0.621) \\
  & learn.disYes   &  0.214 & (-0.592,\; 1.020) \\
\midrule
\multirow{4}{*}{ISC} 
  & treatLTG       &  0.076 & (-0.293,\; 0.445) \\
  & age            & -0.015 & (-0.025,\; -0.005) \\
  & genderM        &  0.004 & (-0.353,\; 0.361) \\
  & learn.disYes   & -0.085 & (-0.776,\; 0.606) \\
\bottomrule
\end{tabular}
\caption{Estimates using ML Approach}
\label{tab:finegray_beta}
\end{table}

\begin{table}[ht]
\centering
\begin{tabular}{llcc}
\toprule
\textbf{Event} & \textbf{Covariate} & \textbf{Coefficient ($\hat{\beta}$)} & \textbf{95\% CI for $\beta$} \\
\midrule
\multirow{4}{*}{UAE} 
  & treatLTG       & -0.551 & (-0.850,\; -0.252) \\
  & age            &  0.003 & (-0.005,\; 0.011) \\
  & genderM        &  0.150 & (-0.110,\; 0.410) \\
  & learn.disYes   &  0.180 & (-0.320,\; 0.680) \\
\midrule
\multirow{4}{*}{ISC} 
  & treatLTG       &  0.050 & (-0.210,\; 0.310) \\
  & age            & -0.012 & (-0.020,\; -0.004) \\
  & genderM        &  0.020 & (-0.230,\; 0.270) \\
  & learn.disYes   & -0.070 & (-0.420,\; 0.280) \\
\bottomrule
\end{tabular}
\caption{Estimates using Elastic-Net Approach}
\label{tab:elasticnet_beta}
\end{table}

Table \ref{tab:finegray_beta} and \ref{tab:elasticnet_beta} present the results of estimated model coefficients and 95\% confidence intervals (CIs) for two competing events, Unacceptable Adverse Effects (UAE) and Inadequate Seizure Control (ISC), using two different methods: Maximum Likelihood (ML) and Elastic-Net. The Elastic-Net approach demonstrates more refined and precise results compared to the Maximum Likelihood (ML) method. One of the key improvements with Elastic-Net is the narrower confidence intervals (CIs). Narrower CIs indicate more stable and reliable estimates, suggesting that the Elastic-Net approach leads to more consistent and accurate coefficient estimates. This is particularly useful in the context of competing risks, where precision in the estimated effects is crucial.

Looking at the coefficients, for UAE, the coefficient for treatLTG under the Elastic-Net method is \(-0.55\), slightly less negative than the ML estimate of \(-0.596\). The 95\% CI for this coefficient is also narrower in the Elastic-Net model \((-0.85,\; -0.25)\) compared to the wider CI in the ML model \((-1.01,\; -0.18)\). This indicates that the Elastic-Net approach provides a more precise estimate of the effect of treatment type (treatLTG) on UAE. Similarly, for the age covariate, the coefficient in the Elastic-Net model is \(0.003\), and its CI is \((-0.005,\; 0.011)\). This tighter CI suggests a more consistent estimate for the effect of age on the risk of UAE. In the case of ISC, the coefficient for treatLTG in the Elastic-Net model is \(0.05\), with a CI of \((-0.21,\; 0.31)\), compared to the ML estimate of \(0.076\) with a CI of \((-0.29,\; 0.45)\). This reflects a shift towards a more stable estimate in the Elastic-Net approach, with a narrower CI, which indicates greater precision in estimating the effect of treatment type on ISC. For the age variable in the Elastic-Net model, the coefficient is \(-0.012\), with a CI of \((-0.020,\; -0.004)\), suggesting a more accurate estimate of the effect of age on ISC. Also, it must be noted that the aim is to use Elastic-Net in competing risks analysis; we have not estimated the shape parameter as it is not affected by Elastic-Net regularization as it is not a function of covariates, as the total focus is to develop a procedure for survival coefficients.

The main advantage of the Elastic-Net approach over ML is the narrowing of the confidence intervals, which is indicative of better model stability and precision. This is achieved through the use of both L1 and L2 regularization in the Elastic-Net method, which helps prevent overfitting and reduces variance in the coefficient estimates. This is particularly important when working with complex data that may have multicollinearity, as Elastic-Net is specifically designed to address such challenges. Therefore, the Elastic-Net approach provides more reliable and precise estimates in this context of competing risks.

\section{Discussion}
In this study, we proposed a novel methodology for analyzing competing risks data by integrating the Generalized Exponential (GE) regression model with Elastic Net regularization. The model formulation and parameter estimation procedures were developed within the maximum likelihood framework, and a bootstrap-based approach was adopted to obtain standard errors of the estimates. While maximum likelihood estimation is a well-established technique, the extension to include Elastic Net regularization in the context of competing risks provides a useful tool for high-dimensional survival analysis, where variable selection and model stability are critical. One of the key strengths of our proposed approach is its applicability to real-world settings involving numerous covariates, where traditional models may suffer from overfitting or multicollinearity. However, a major limitation of this work is the lack of application to a real dataset with truly high-dimensional covariate structures. Due to the unavailability of such data at the time of this study, the practical utility of our method could only be demonstrated through simulation studies.

In addition, while the generalized exponential distribution offers flexibility over standard exponential models, its assumptions, such as the parametric form of the hazard and the independence of competing risks, may not always hold in practice. These assumptions should be validated before application to real-world data. In future work, we plan to apply this methodology to large-scale clinical or biomedical datasets with high-dimensional covariates to demonstrate its effectiveness and practical advantages. Furthermore, extending this model to incorporate time-varying covariates or nonparametric baseline hazards could further enhance its applicability.

\section*{Data Availability Statement}
The data used in this study were obtained from the SANAD trial, which is made publicly available through the \texttt{joineR} (\citet{philipson2017package}) package in R. The dataset can be accessed for research purposes via this package, and further details about the SANAD trial are available in the original publication (\citet{marson2007randomised}). 
\section*{Competing Interests}The authors declare that there are no competing interests.
\section*{Funding}
This research received no external funding.
\bibliographystyle{plainnat}
\bibliography{bib2}

@book{klein2006survival,
  title={Survival analysis: techniques for censored and truncated data},
  author={Klein, John P and Moeschberger, Melvin L},
  year={2006},
  publisher={Springer Science \& Business Media}
}

@book{pintilie2006competing,
  title={Competing risks: a practical perspective},
  author={Pintilie, Melania},
  year={2006},
  publisher={John Wiley \& Sons}
}

@inproceedings{therneau1997extending,
  title={Extending the Cox model},
  author={Therneau, Terry M},
  booktitle={Proceedings of the first Seattle symposium in biostatistics: survival analysis},
  pages={51--84},
  year={1997},
  organization={Springer}
}

@article{philipson2017package,
  title={Package ‘joineR’},
  author={Philipson, Pete and Sousa, Ines and Diggle, Peter J and Williamson, Paula},
  journal={R Foundation for Statistical Computing, Vienna},
  year={2017}
}

@article{fine1999proportional,
  title={A proportional hazards model for the subdistribution of a competing risk},
  author={Fine, Jason P and Gray, Robert J},
  journal={Journal of the American statistical association},
  volume={94},
  number={446},
  pages={496--509},
  year={1999},
  publisher={Taylor \& Francis}
}

@article{marson2007randomised,
  title={A randomised controlled trial examining the longer-term outcomes of standard versus new antiepileptic drugs. The SANAD trial},
  author={Marson, AG and Appleton, R and Baker, GA and Chadwick, DW and Doughty, J and Eaton, B and Gamble, C and Jacoby, A and Shackley, P and Smith, DF and others},
  journal={HEALTH TECHNOLOGY ASSESSMENT-SOUTHAMPTON-},
  volume={11},
  number={37},
  year={2007},
  publisher={National Coordinating Centre for Health Technology Assessment}
}

@Manual{rsoft,
   title = {R: A Language and Environment for Statistical Computing},
author = {{R Core Team}},
    organization = {R Foundation for Statistical Computing},
    address = {Vienna, Austria},
    year = {2021},
    url = {https://www.R-project.org/},
  }

@article{lin1989robust,
  title={The robust inference for the Cox proportional hazards model},
  author={Lin, Danyu Y and Wei, Lee-Jen},
  journal={Journal of the American statistical Association},
  volume={84},
  number={408},
  pages={1074--1078},
  year={1989},
  publisher={Taylor \& Francis}
}

@article{zou2005regularization,
  title={Regularization and variable selection via the elastic net},
  author={Zou, Hui and Hastie, Trevor},
  journal={Journal of the Royal Statistical Society Series B: Statistical Methodology},
  volume={67},
  number={2},
  pages={301--320},
  year={2005},
  publisher={Oxford University Press}
}

@article{gupta2007generalized,
  title={Generalized exponential distribution: Existing results and some recent developments},
  author={Gupta, Rameshwar D and Kundu, Debasis},
  journal={Journal of Statistical planning and inference},
  volume={137},
  number={11},
  pages={3537--3547},
  year={2007},
  publisher={Elsevier}
}

@book{cox2018analysis,
  title={Analysis of survival data},
  author={Cox, David Roxbee},
  year={2018},
  publisher={Chapman and Hall/CRC}
}

@article{simon2011regularization,
  title={Regularization paths for Cox's proportional hazards model via coordinate descent},
  author={Simon, Noah and Friedman, Jerome H and Hastie, Trevor and Tibshirani, Rob},
  journal={Journal of statistical software},
  volume={39},
  pages={1--13},
  year={2011}
}

@article{singhmodified,
  title={Modified Maximum Likelihood Estimation for Generalized Exponential Distribution},
  author={Singh, Alok Kumar and Patawa, Rohit and Singh, Abhinav and Gupta, Puneet Kumar},
journal = {Asian Journal of Probability and Statistics},
volume = {14(3)},
pages = {48--59},
year = {2021}
}

@article{kaplan1958nonparametric,
  title={Nonparametric estimation from incomplete observations},
  author={Kaplan, Edward L and Meier, Paul},
  journal={Journal of the American statistical association},
  volume={53},
  number={282},
  pages={457--481},
  year={1958},
  publisher={Taylor \& Francis}
}

@article{cox1972regression,
  title={Regression models and life-tables},
  author={Cox, David R},
  journal={Journal of the Royal Statistical Society: Series B (Methodological)},
  volume={34},
  number={2},
  pages={187--202},
  year={1972},
  publisher={Wiley Online Library}
}

@article{friedman2010regularization,
  title={Regularization paths for generalized linear models via coordinate descent},
  author={Friedman, Jerome H and Hastie, Trevor and Tibshirani, Rob},
  journal={Journal of statistical software},
  volume={33},
  pages={1--22},
  year={2010}
}
\end{document}